\documentclass[prb,twocolumn,showpacs,superscriptaddress,floatfix]{revtex4}

\usepackage{graphicx}
\usepackage{amssymb}

\begin{document}

\title{Phase diagram of the one dimensional Hubbard-Holstein Model at 1/2 and 
1/4 filling}
\author{R.P. Hardikar and R.T. Clay}
\email[]{r.t.clay@msstate.edu}
\affiliation{Department of Physics and Astronomy and $\rm{HPC^{2}}$ 
Center for Computational Sciences, Mississippi State University, 
Mississippi State MS 39762}

\date{\today}

\begin{abstract}
The Hubbard-Holstein model is one of the simplest to incorporate both
electron-electron and electron-phonon interactions. In one dimension
at half filling the Holstein electron-phonon coupling promotes onsite
pairs of electrons and a Peierls charge density wave while the Hubbard
onsite Coulomb repulsion $U$ promotes antiferromagnetic correlations
and a Mott insulating state.  Recent numerical studies have found a
possible third intermediate phase between Peierls and Mott states.
From direct calculations of charge and spin susceptibilities, we show
that (i) As the electron-phonon coupling is increased, first a spin
gap opens, followed by the Peierls transition.  Between these two
transitions the metallic intermediate phase has a spin gap, no charge
gap, and properties similar to the negative$-U$ Hubbard model. (ii)
The transitions between Mott/intermediate and intermediate/Peierls
states are of the Kosterlitz-Thouless form. (iii) For larger $U$ the
two transitions merge at a tritical point into a single first order
Mott/Peierls transition.  In addition we show that an intermediate
phase also occurs in the quarter-filled model.
\end{abstract}

\maketitle

\section{Introduction}
In crystalline materials where one or more of the building blocks of
the crystal structure is a large molecule the vibrational properties
of the molecules often have large effects on the overall electronic
properties of the material.  One large family of such molecular
crystalline materials are the organic conductors and
superconductors\cite{Ishiguro}. While some molecular crystals such as
the fullerene superconductors \cite{Gunnarsson97a} have a
three-dimensional crystal structure, many other examples are either
quasi-one or quasi-two dimensional, {\it i.e.}  charge transport is
restricted in certain directions due to anisotropic crystal
structure. In addition to strong electron-phonon (e-ph) coupling to
the molecular vibrations, electron-electron (e-e) interactions are
often important in low dimensional materials. In this paper we
present numerical calculations of the phase diagram for one of the
simplest possible many-body models incorporating both these effects,
the Hubbard-Holstein model (HHM) in one dimension (1D). In the HHM,
internal (intra-molecular) molecular vibrations are coupled to the
local charge density of the electrons\cite{Holstein59a}. The electrons
further interact with other electrons with an onsite Coulomb repulsion
when two electrons occupy the same orbital \cite{Hubbard63a}.
Surprisingly complex effects result from this simple model due to the
presence of both e-e and e-ph interactions.

The 1D HHM Hamiltonian we consider is
\begin{eqnarray}
  H &=& -t \sum_{j,\sigma}\left(c^{\dagger}_{j+1,\sigma} c_{j,\sigma}
  + h.c.\right) + U \sum_{j} n_{j,\uparrow}n_{j,\downarrow} \nonumber\\
  &+ & g \sum_{j,\sigma}\left(a^{\dagger}_{j} + a_{j} \right)
  n_{j,\sigma} + \omega \sum_{j} a^{\dagger}_{j}a_{j}, \label{ham}
\end{eqnarray}
where $c^{\dagger}_{j,\sigma} \left(c_{j,\sigma}\right)$ are creation
(annihilation) operators for electrons on site $j$ with spin $\sigma$,
$a^{\dagger}_{j} \left(a_{j}\right)$ are bosonic creation
(annihilation) operators for phonons at site $j$, and the electron
number operator $n_{j,\sigma} = c^\dagger_{j,\sigma}c_{j,\sigma}$. $U$
is the Hubbard onsite e-e interaction energy, $\omega$ is the
dispersionless phonon frequency, and $g$ is the e-ph coupling
constant. All energies in this paper will be given in units of $t$,
the electron hopping integral.

We will concentrate primarily on Eq.\ref{ham} in the half-filled band
limit (one electron per lattice site), but also discuss briefly the
quarter-filled band (one electron per two lattice sites). The effect of
e-ph interactions on a half-filled 1D metal is well known: for
inter-molecular phonons corresponding to the relative motion of
adjacent molecules in the crystal, the 1D lattice dimerizes with
alternating strong and weak bonds.  In this bond-order wave (BOW)
state the expectation value of the electron hopping between adjacent
sites alternates between strong and weak values.  The dimerized chain
then has a gap at the Fermi level and an insulating ground state
\cite{Peierls}. This Peierls state has both charge and spin gaps, and
a bond modulation at 2k$_F$ ($q=\pi$) at half filling. For
Holstein-type phonons that couple to the local charge density a
similar Peierls state occurs, but instead of bond deformation the
local charge density is modulated in a charge density wave (CDW)
ground state. The CDW Peierls state at half filling has alternating
large and small charge densities again with periodicity
2k$_F$. Similarly, the effect of the Hubbard onsite interaction in 1D
is well known: for any $U>0$ at half filling, the ground state is an
insulator \cite{Lieb68a}. Anti-ferromagnetic (AFM) spin correlations
are present in this Mott insulating state, although no long-range
antiferromagnetic order is possible in 1D.  At half filling, the 2k$_F$
CDW cannot coexist with 2k$_F$ AFM correlations and hence the Peierls
and AFM states are competing.

Numerous previous studies have examined HHM within various
approximations and analytic or numerical techniques.  In the limit
$\omega\rightarrow\infty$ one can integrate out the phonons leaving an
effective $U$ composed of the sum of the Hubbard $U$ and the effective
phonon interaction, $U_{\rm{eff}}=U-2g^2/\omega$. For $U_{\rm{eff}}>0$
one expects the Mott state, while for $U_{\rm{eff}}<0$ one expects the
Peierls state \cite{Hirsch82a,Hirsch83a}. If the phonons are treated
in the classical (adiabatic $\omega\rightarrow 0$) limit, one expects
Peierls order for any $g>0$ at $U=0$. However, it was shown in the
spinless model (Eq.~\ref{ham} with a single species of fermion) that
quantum fluctuations of the phonon field lead to a finite e-ph
coupling $g_c$ before the Peierls state is formed at $U=0$
\cite{Hirsch82a,Hirsch83a,Bursill98a}. The model with spin
(Eq.~\ref{ham}) has since been shown to also require a finite e-ph
coupling for the Peierls transition \cite{Wu95a,Jeckelmann99a}.

In addition to studies of the 1D model several recent studies have
been performed on the Holstein and HHM in the limit of infinite
dimensions ($d=\infty$) using dynamical mean-field theory (DMFT) and
related methods \cite{Meyer02a,Capone04a,Koller04a}. In the $d=\infty$
model the system is metallic at $U=0$ also for $g$ less than a finite
value. However, an important distinction between $d=\infty$ results
and those presented here is that at $d=\infty$, the Mott insulating
transition occurs at a finite value of $U$, $U\agt 6t$, while at $d=1$ it
occurs for $U>0$. Some similarities are found with our results, in
particular that there is a deviation in the critical coupling for the
Peierls transition from $U_{\rm{eff}}=0$ at small $U$
\cite{Koller04a}.

Given that in the half-filled 1D HHM at $U=0$ the ground state is
metallic (no charge gap and a finite Drude weight) for a finite value
of $g$, it was proposed that this metallic phase continues to exist
{\it between} the Peierls and Mott insulating phases for $U>0$
\cite{Takada03a}. Subsequent numerical calculations confirmed that a
metallic phase exists for both $U=0$ and finite $U$ \cite{Clay05b}. In
this paper, we present more detailed numerical results and analysis of
the phase diagram. We confirm the intermediate phase using a different
and more direct order parameter, and present more detailed finite-size
scaling of the quantum phase transitions.  From the finite-size
dependence, we determine that the two transitions (Mott/intermediate
and intermediate/Peierls) are of the Kosterlitz-Thouless (KT) type. We
find that for larger $U$, the two transitions merge into a single
first-order Mott/Peierls transition.  In our revised analysis, we find
that the apparent presence of the Luttinger Liquid (LL) exponent
$K_\rho>1$ \cite{Clay05b} {\it does not} imply dominant
superconducting pairing correlations, but is more likely a finite-size
effect.  We present the phase diagram for three different phonon
frequencies.  We further show that at quarter filling a similar
intermediate phase occurs.

The outline of the paper follows. We first give some details of the
numerical method we used. Turning to our results, we discuss the $U=0$
case and then move on to finite $U$ and the quarter-filled band.  Finally,
we conclude with a discussion of our data and their relation to other
theoretical results, as well as unanswered questions for further
study.

\section{Method}

We use the Stochastic Series Expansion (SSE) quantum Monte Carlo (QMC)
method
\cite{Sandvik92a,Sandvik99a,Sandvik99b,Sengupta03a,Syljuasen02a}.  SSE
provides statistically exact results (no Trotter discretization of
imaginary time is used) and has been adapted for many different
quantum lattice models.  Although this method has been described in
detail elsewhere, we briefly describe here our treatment of the
Holstein phonon interaction.

 In SSE, the partition function $Z=Tr\{e^{-\beta H}\}$ is expanded in
terms of a series of sequences $S_L$ of operators
$H_{a_i,b_i}$:
\begin{equation}
Z=\sum_\alpha \sum_{S_L} \frac{\beta^n(L-n)!}{L!}
\langle\alpha|\prod_{i=1}^L H_{a_i,b_i} | \alpha\rangle
\label{eqn:sse}
\end{equation}
In Eq.~\ref{eqn:sse}, $n$ is the length (number of operators) of each
sequence, $L$ the maximum allowed sequence length, and $\beta$ is the
inverse temperature and $|\alpha\rangle$ is a basis state, here a
direct product of electron and phonon configurations.  In order to
obtain the ground-state phase diagram, all results presented here used
$\beta/t \ge 2N$, where $N$ is the number of lattice sites.  The
operators $H_{a_i,b_i}$ define the Hamiltonian, and have type ($a_i$)
and bond ($b_i$) indices with $i$ indicating their position within the
sequence $S_L$. For the 1D Hubbard model (Eq.~\ref{ham} with
$g=\omega=0$), we have three different operators representing the
diagonal interaction and electron hopping for both
spins\cite{Sandvik92a}:
\begin{eqnarray}
H_{1,j} &=& C -\frac{U}{2}[(n_{\uparrow,j}-\frac{1}{2})
(n_{\downarrow,j}-\frac{1}{2}) \nonumber \\
&+&(n_{\uparrow,j+1}-\frac{1}{2})(n_{\downarrow,j+1}-\frac{1}{2})] \nonumber \\
&+&\mu(2-n_j - n_{j+1}) \label{eqn:h1}  \\
H_{2,j} &=& c_{j+1,\uparrow}^{\dag}c_{j,\uparrow} + h.c. \label{eqn:h2} \\
H_{3,j} &=& c_{j+1,\downarrow}^{\dag}c_{j,\downarrow} + h.c.
\label{tuvops}
\end{eqnarray}
Here $j$ labels the first site of the bond the operator acts on. $\mu$
is the chemical potential, written here so that $\mu=0$ corresponds to
half filling. $C$ is a constant chosen so that the expectation value of
$H_{1,j}$ is always positive definite.  In addition to the operators
of Equations \ref{eqn:h1}---\ref{tuvops}, a null operator $H_0$ is
used as a place-holder in the sequence expansion.  We represent the
phonons in the phonon-number basis and add the following operators for
the e-ph interactions and phonon diagonal energy:
\begin{eqnarray}
H^L_{4,j} &=& ga_j^{\dag}n_j \label{eqn:h4} \\
H^R_{4,j} &=& ga_{j+1}^{\dag}n_{j+1} \\
H^L_{5,j} &=& ga_j n_j \\
H^R_{5,j} &=& ga_{j+1} n_{j+1} \label{eqn:h5} \\
H_{6,j} &=& \omega(N_{\rm{p}}- a_j^\dagger a_j) \label{eqn:h6}
\end{eqnarray}
Additionally, for the HHM, $\mu$ in Eq.~\ref{eqn:h1} should be
replaced by (2g$^2/\omega+\mu$).  Since the Holstein interaction
couples the electron density on a single site while the SSE operators
typically act on bonds composed of two sites, we define two different
phonon operators acting on phonon numbers on the left or right of the
bond. These have superscripts ``L'' and ``R'' respectively.  The
diagonal operator $H_{6,j}$ also acts on a single site $j$.  $N_p$ is
a cutoff in the maximum number of phonons per site.  We discuss
further below the choice of this cutoff, but in practice it can be
chosen large enough so as to not affect the accuracy of the method.

The Monte Carlo updating is composed of an update for the electrons
followed by an update for the phonons. The electron update consists of
an update changing the number of diagonal $H_{1,j}$ operators in the
sequence, followed by a loop update that exchanges diagonal and
off-diagonal operators. For the electrons we use the directed loop
algorithm \cite{Syljuasen02a}. We note that the operators
Eq.~\ref{eqn:h4} through Eq.~\ref{eqn:h6} are not changed during the
electron loop update.  The phonon updating also consists of two parts,
first a diagonal update changing numbers of $H_{6,j}$ operators, and
second an off-diagonal update exchanging $H_1$, $H_4$, and $H_5$
operators. In the diagonal phonon update, $H_0$ operators are
interchanged with $H_{6,j}$ operators with the following Metropolis
algorithm probabilities ($N_H$ is the total number of non-$H_0$
operators present in the sequence):
\begin{eqnarray}
P_{0\rightarrow 6}&=& \frac{N\beta \omega(N_{\rm{p}}-\langle a^\dagger_j
a_j\rangle)}{L-N_H} \\
P_{6\rightarrow 0}&=& \frac{L-N_H+1}{N\beta \omega(N_{\rm{p}}-\langle 
a^\dagger_j a_j\rangle)}
\end{eqnarray}

The phonon update for off-diagonal operators is similar to the
technique described in Reference \onlinecite{Sandvik97b}.  For each
site in the system, a {\it subsequence} is constructed which is a
subset of the operators in $S_L$.  The subsequence consists of only
the operators $H_{1,m}$, $H_{4,m}$, and $H_{5,m}$ {\it which act on
phonons at a particular site $m$}.  Within the subsequence, adjacent
pairs of operators are then selected at random and changed with a
Metropolis probability. The pair substitutions that change the phonon
number are (omitting the site index $m$ as all apply to the same
site):
\begin{eqnarray}
(H_1,H_1) & \rightarrow & (H_4,H_5), (H_5,H_4)  \\
(H_4,H_5) & \rightarrow & (H_1,H_1) \\
(H_5,H_4) & \rightarrow & (H_1,H_1) 
\end{eqnarray}
In addition, pair substitutions are attempted that swap the order in
the subsequence of the two operators.  When two different pairs may be
substituted the substitution made is chosen randomly.  Note that the
$L$ and $R$ indices in Eq.~\ref{eqn:h4} through Eq.~\ref{eqn:h5} are
not needed during the pair updating, but updates involving the $H_1$
operators must be canceled with 50\% probability (for each $H_1$
operator in the pair).  If a $H_1$ operator changes into a phonon
operator as a result of the update, a $L$ or $R$ index is assigned
when the subsequence update is completed and merged into $S_L$.  The
Metropolis substitution probabilities depend on phonon as well as
diagonal electron matrix elements variables, the e-ph coupling
constant $g$, and the number $N_{\rm{d}}$ of diagonal phonon operators
($H_{6,j}$) that are present between the two operators of the pair.
$N_{\rm{d}}$ may be stored when the subsequence is constructed.  For
example, in terms of just the change in the phonon part of the
operator,
\begin{eqnarray}
P[(H_1,H_1) \rightarrow (H_5,H_4)] &=& Rng^2 \left[\frac{N_{\rm{p}} - 
n + 1}{N_{\rm{p}} - n}\right]^{N_{\rm{d}}} \label{pair1}\\
P[(H_5,H_4) \rightarrow (H_1,H_1)] &=& \frac{R}{ng^2}
 \left[\frac{N_{\rm{p}} - 
n}{N_{\rm{p}} - n + 1}\right]^{N_{\rm{d}}}   \label{pair2}
\end{eqnarray}
where $n$ is the number of phonons present in the sequence position
just before the operator pair.  $R$ in Eq.~\ref{pair1} and
Eq.~\ref{pair2} is the ratio of diagonal matrix elements from the
electronic Hamiltonian.  In practice, the number of pair substitutions
performed is chosen to be approximately the same as the number of
operators in the subsequence.

 We use standard methods to calculate various observables within our
SSE code \cite{Sandvik92a}. To determine phase boundaries of the model
we primarily use the charge and spin susceptibilities at wavevector
$q$ given by
\begin{eqnarray}
O^\pm_j & = & n_{j,\uparrow} \pm n_{j,\downarrow}  \\
  \chi_{\rho,\sigma}(q) &=& \frac{1}{N} \sum_{j,k} e^{iq(j-k)} \int_{0}^{\beta} d\tau
 \langle O^\pm_j(\tau)O^\pm_k(0)\rangle
\label{eqn:susceptibility}
\end{eqnarray}
In Eq.~\ref{eqn:susceptibility}
the charge susceptibility $\chi_\rho(q)$ (spin susceptibility $\chi_\sigma(q)$) 
corresponds to the  $+$ ($-$) sign.
Similarly, we also use the static structure factors, $S_\rho(q)$ and
$S_\sigma(q)$:
\begin{equation}
  S_{\rho,\sigma} = \frac{1}{N} \sum_{j,k} e^{iq(j-k)}
  \langle O^\pm_j O^\pm_k\rangle
\label{eqn:st-fac} 
\end{equation}

The effective low-energy properties of many interacting 1D models can
be understood in terms of a LL picture, and the
asymptotic properties of the system described by a small number of
parameters \cite{Schulz87a,Voit95a}.  In particular, the asymptotic
\begin{figure}[tb]
\centerline{\resizebox{3.0in}{!}
{\includegraphics{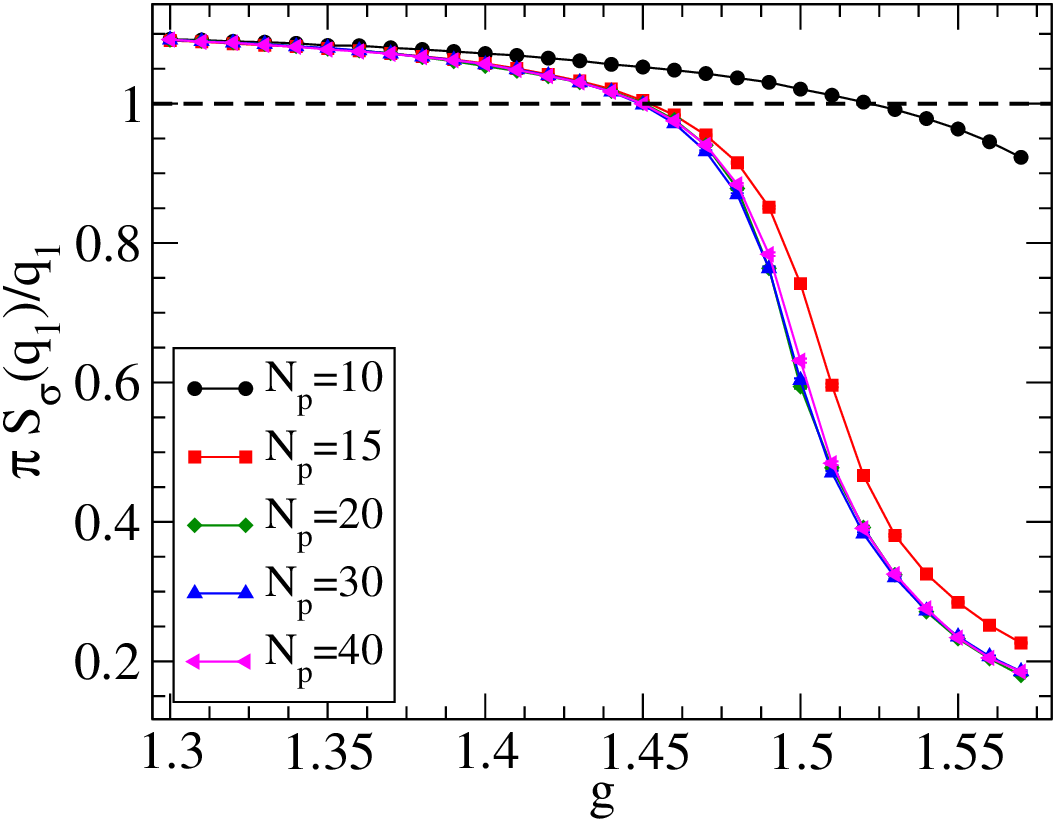}}}
\caption{(color online) Slope of the spin structure factor at
wavevector $q_1=2\pi/N$ versus e-ph coupling $g$ for a 16 site
half-filled system with $U=4$ and $\omega=1$.  $\pi S_\sigma(q_1)/q_1$
crossing one indicates the opening a spin gap. Different symbols show
the convergence with increasing phonon cutoff $N_p$.}
\label{fig-npmax}
\end{figure}
decay of correlation functions can be related to the correlation
exponents $K_\rho$ and $K_\sigma$ for charge and spin respectively.
In the long wavelength limit, these exponents may be calculated from
the slope of the structure factors:
\begin{equation}
K_{\rho,\sigma}= \pi \lim_{q\rightarrow 0} S_{\rho,\sigma}(q)/q
\label{krho}
\end{equation}
In practice, one uses the behavior of $\pi S(q)/q$ at the smallest
available $q$ for the periodic ring, $q_1=2\pi/N$. With proper
finite-size scaling in $N$, this gives the Luttinger liquid exponent
for the system \cite{Clay99a}.  Based on calculations of acoustic
phonons coupled to 1D electrons it has been suggested that the
expected relationship of $K_\rho$ to the correlation functions must be
modified in the presence of phonon interactions with retardation
\cite{Loss94a,Tezuka05a,Tam06a}.  We will discuss this further in
Section \ref{sect:krho} below. However, we note that the
interpretation of $K_\sigma$ is {\it not} modified in the presence of
phonon retardation effects since spin-rotation symmetry is preserved
in the HHM.  $K_\sigma$ is expected to be exactly equal to one unless
a spin gap is present, and the condition that $\pi S_\sigma(q_1)/q_1$
decreases below one is a sensitive indicator for the opening of a spin
gap (see Fig.~\ref{fig-npmax})\cite{Sengupta03a}.  We find that
finite-size effects in determining the phase boundaries using
Eq.~\ref{krho} are worse than when using the susceptibilities,
Eq.~\ref{eqn:susceptibility}, due to the necessity of taking the limit
$q_1\rightarrow 0$ in Eq.~\ref{krho}.  Therefore we will primarily use
the susceptibilities in order to determine the phase diagram
boundaries.

We choose the phonon cutoff $N_{\rm{p}}$ such that phonon occupation
numbers during the simulation never reach within some fraction
($\sim$20\%) of the cutoff, similar to the method in which the maximum
sequence length $L$ is set self-consistently in SSE simulations.  We
have verified that our results are converged with respect to
$N_{\rm{p}}$.  Typical variation with $N_{\rm{p}}$ is shown in
Fig.~\ref{fig-npmax} for a 16 site system with $U=4$ and $\omega=1$.
We find that choosing $N_p$ too small can have a noticeable effect on
the critical coupling for transitions, and especially on quantities
measured in the Peierls phase.

\begin{figure}[tb]
\centerline{\resizebox{3.0in}{!}
{\includegraphics{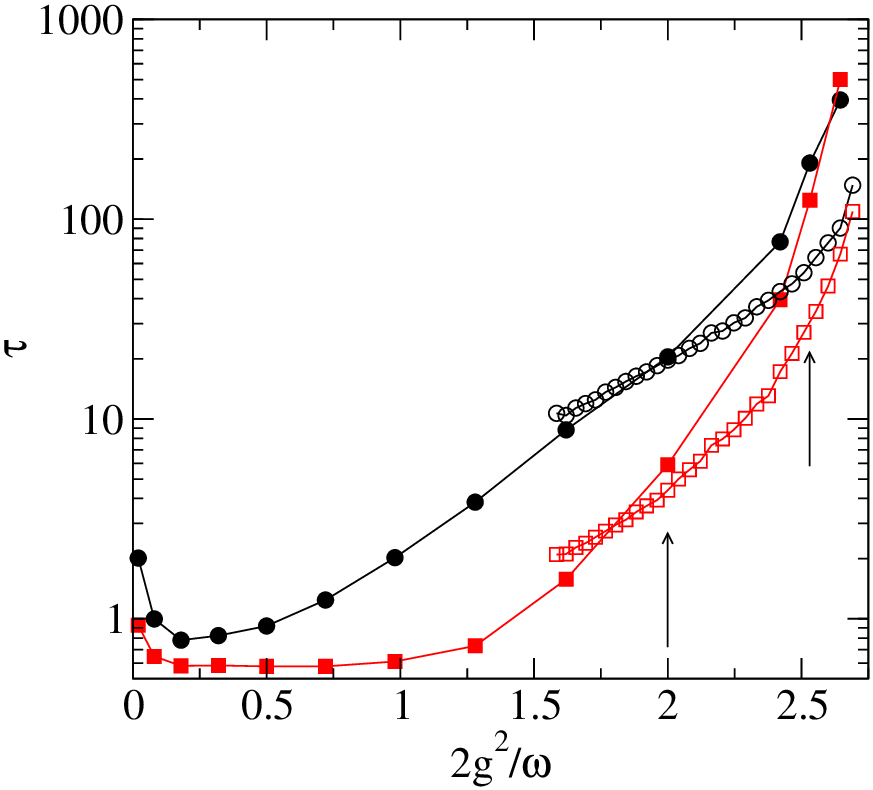}}}
\caption{(color online) Integrated autocorrelation time for $N=16$,
$U=2$, $\omega=1$, $\beta=32$, $N_p=30$, as a function of e-ph
coupling.  Filled circles (squares) are the autocorrelation time for the charge
(spin) structure factor $S(q_1)$ at $q_1=2\pi/N$.  Open symbols are
for the same observables, but calculated using quantum parallel
tempering. Arrows indicate the location of the two transitions (see Section
\ref{sect:half}).  We find that parallel tempering significantly
reduces the autocorrelation time near the transitions.}
\label{fig-ac}
\end{figure}

Autocorrelation time $\tau$ is an important measure of the overall
efficiency of a Monte Carlo method. Correlations between measurements
are typically found to decay as $\sim e^{-t/\tau}$, where $t$ is in
units of Monte Carlo time corresponding to the number of updating
steps completed. If measurements are correlated the estimated
statistical error must be increased. In general, it is found that near
quantum phase transitions, $\tau$ often increases steeply, making
calculations near phase boundaries difficult or impossible.  One tool
available to improve QMC calculations near phase boundaries is quantum
parallel tempering \cite{Sengupta02a}. In this technique, separate
processors on a parallel computer have slightly different parameter
values. Periodically a Metropolis move is attempted to switch the
configuration between adjacent processors. These moves help to prevent
the algorithm from getting ``stuck'' in one configuration, and
consequently reduce the autocorrelation time.  In Fig.~\ref{fig-ac} we
show the integrated autocorrelation time for long wavelength structure
factor measurements (Eq.~\ref{krho}), defined as in Reference
\onlinecite{Syljuasen02a}. Our definition of one Monte Carlo step is
similar to Reference \onlinecite{Syljuasen02a}, with an average $2N_H$
loop vertices visited in the electronic loop update.  As expected we
find that $\tau$ increases greatly near the Peierls transition. We
also find that parallel tempering decreases the autocorrelation time
significantly and is essential to obtain reliable results near the
Peierls transition.

\section{Results at Half-filling}
\label{sect:half}

We first present our results for the  half-filled band, first in the
case $U=0$ and then for finite $U$.
\begin{figure}[tb]
\centerline{\resizebox{3.2in}{!}
{\includegraphics{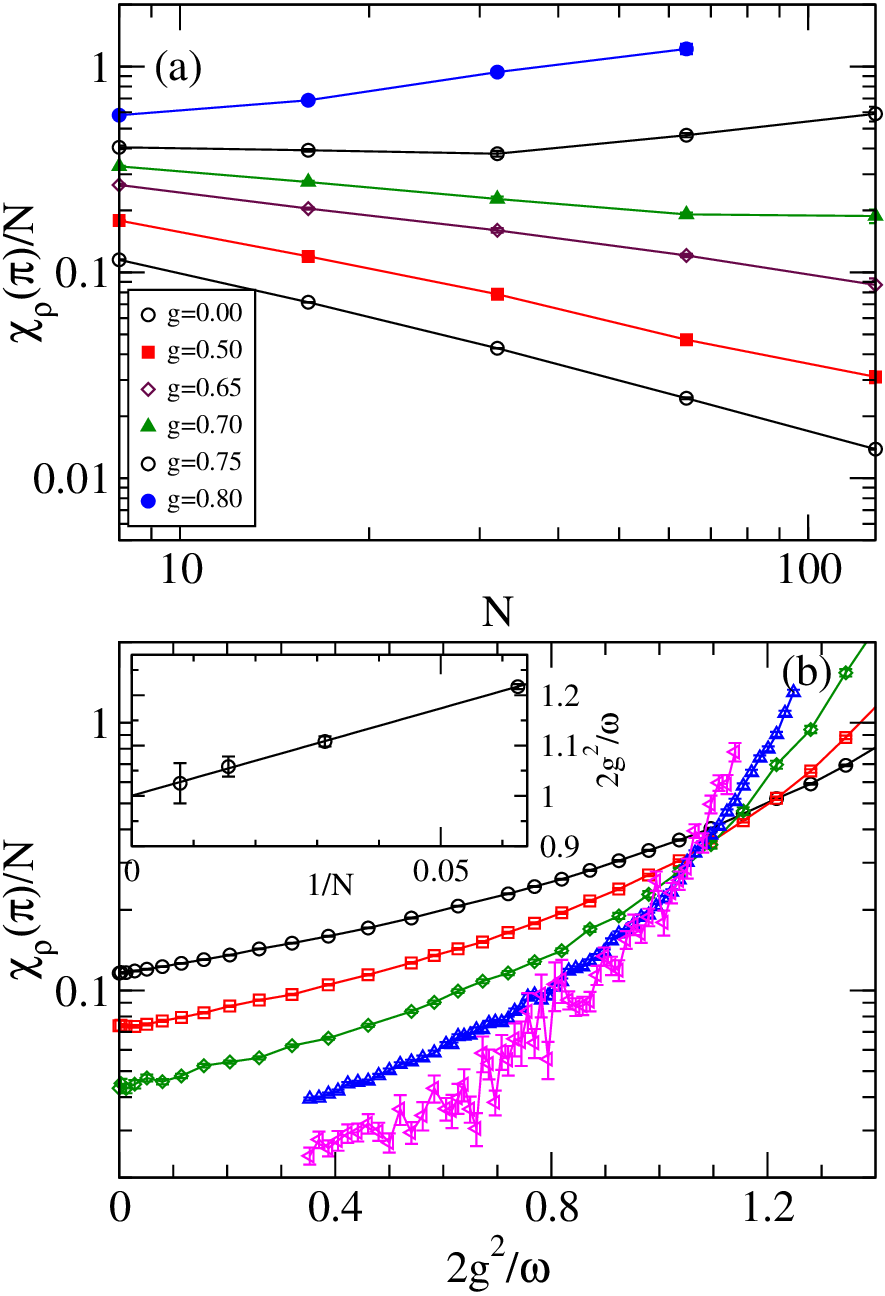}}}
\caption{(color online) Finite-size scaling of the $q=\pi$ charge
susceptibility for $U=0$ and $\omega=1$ at half filling. Data are for
system sizes up to $N=128$ sites. In (a) we plot $\chi_\rho(\pi)/N$
versus $N$.  At critical coupling $\chi_\rho(\pi)/N$ approaches a
constant for large $N$.  Note that the $g=0$ curve corresponds to free
fermions (no phonons). In (b) $\chi_\rho(\pi)/N$ is plotted versus the
effective e-ph coupling $2g/\omega$, for system sizes $N=8$ (open
circles), 16, 32, 64, and 128. The inset shows finite-size scaling of
the transition point obtained by plotting the value of 2g$^2/\omega$
where $\chi_\rho(\pi)/N$ for system size $N$ exceeds the
susceptibility for system size $N/2$. Line in inset is a linear fit.
We estimate the critical coupling as $2g_{c2}^2/\omega\approx 1.00$
($g_{c2}\approx 0.71$). }
\label{u0w1_chi_rho}
\end{figure}

\subsection{$U=0$ : the Peierls transition}
\label{sect:u0}

Eq.~\ref{ham} has been studied in great detail for the case of
$U=0$. One of the key questions is whether the transition to the
Peierls state occurs for finite critical coupling or for any value of
$g>0$. The transition occurring at finite $g$ is expected to be of the
KT type\cite{Hirsch82a,Fradkin83a}.  KT transitions at finite phonon
coupling have been found in a number of 1D phonon-coupled models
including the spinless Holstein model (Eq.~\ref{ham} with only one
species of fermion) \cite{Hirsch82a,Bursill98a}, the XY model coupled
to dispersionless phonons \cite{Caron96a}, the Heisenberg model
coupled to dispersionless phonons \cite{Sandvik99b}, and the extended
Peierls-Hubbard model coupled to dispersionless bond phonons
\cite{Sengupta03a}. We confirm that indeed a finite critical coupling
exists and show that the finite-size scaling of the observables is
consistent with a KT transition.

  A KT quantum phase transition is difficult to detect because the gap
opens exponentially slowly. For Holstein-type phonons that couple to
the local electron density, the appropriate order parameter for the
transition is the 2k$_{\rm{F}}$ charge susceptibility.  The critical
coupling (we will denote the critical $g$ for the Peierls transition
as $g_{c2}$) may be determined from the finite-size scaling of the
2k$_{\rm{F}}$ charge susceptibility, $\chi_\rho(\pi)$.
\begin{figure}[tb]
\centerline{\resizebox{3.2in}{!}
{\includegraphics{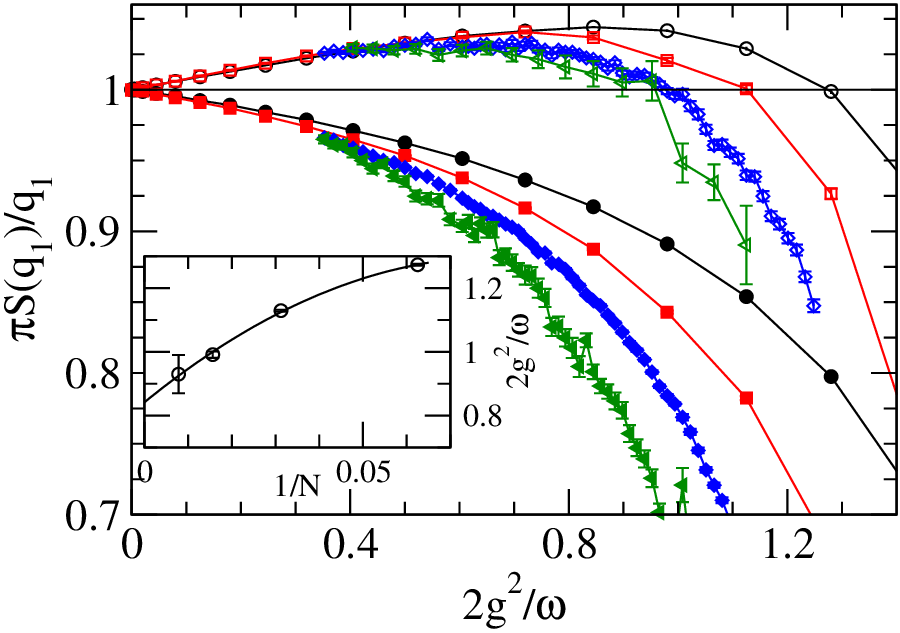}}}
\caption{(color online) Long-wavelength spin and charge structure
factor slopes for $U=0$ and $\omega=1$ at half filling. Open (filled)
symbols are for charge (spin).  Data are for system sizes of $N=16$,
32, 64, and 128 sites. For any $g>0$ $\pi S_\sigma(q_1)/q_1$ is less
than 1 indicating the presence of a spin gap.  The inset shows the
finite-size scaling of the point where $S_\rho(q_1)/q_1$ crosses one,
with the line a fit to a quadratic.  We estimate the critical coupling
as $2g_{c2}^2/\omega\approx 0.85$.  The appearance of these data for $g<g_{c2}$
are similar to those for the negative $U$ Hubbard model
(Fig.~\ref{1dhubbard} for $U<0$). The interpretation of these data is
discussed in Section \ref{sect:krho}.}
\label{u0w1_ll}
\end{figure}
$\chi_\rho(\pi)/N$ should approach zero logarithmically below $g_{c2}$
and should diverge above $g_{c2}$. Exactly at $g=g_{c2}$, log
corrections vanish and $\chi_\rho(\pi)/N$ should approach a constant
value with increasing $N$.  Our SSE results confirm that
$\chi_\rho(\pi)$ does scale in this manner.  In
Fig.~\ref{u0w1_chi_rho}(a) we show charge susceptibility data for
$U=0$ and $\omega=1$, which is consistent with a KT transition at
$g_{c2}\approx 0.7$. We see a clear decrease of $\chi_\rho(\pi)/N$
with system size below the transition and a clear increase above the
transition. Plotted as a function of effective e-ph coupling
$2g^2/\omega$ (Fig.~\ref{u0w1_chi_rho}(b)), $\chi_\rho(\pi)/N$ for
different $N$ cross at the transition.  In the
Fig.~\ref{u0w1_chi_rho}(b) we show a finite-size scaling of the
transition point obtained by plotting value of 2g$^2/\omega$ where the
susceptibility curve for $N$ sites intersects the data for $N/2$
sites. We find that these intersection points are well fit to a linear
dependence in $1/N$, giving $2g_{c2}^2/\omega$=1.00 for $U=0$.

In Fig.~\ref{u0w1_ll} we show for comparison the long-wavelength
charge and spin structure factor slopes, Eq.~\ref{krho}, which are
estimates for the LL exponents $K_\rho$ and $K_\sigma$.  For any $g>0$
$K_\sigma$ is less than 1 and decreases with increasing chain length,
indicating a spin gap.  Furthermore, in the spin susceptibility (not
shown here), we find no sign of any transition at the critical
coupling where $\chi_\rho(\pi)/N$ diverges.  We denote the critical
coupling for the spin gap opening as $g_{c1}$.  Hence we conclude that
a spin gap is present for {\it any} $g>g_{c1}=0$ when $U=0$, but a
charge gap is only present for $g>g_{c2}$.  In the inset of
Fig.~\ref{u0w1_ll} we show the finite-size scaling of point where
$K_\rho=1$ ($2g_{c2}^2/\omega$). We discuss further the $K_\rho$ data
in Section \ref{sect:krho} and the apparent small discrepancy between
$g_{c2}$ determined from susceptibility versus $K_\rho$ data.

\subsection{$U>0$ : Intermediate phase}
\label{sect:finiteu}

We next consider the case with $U>0$ at half filling.  To avoid any
possible difficulties of interpreting numerical estimates for
$K_\rho$, we determine all phase boundaries {\it directly from
susceptibilities} and $K_\sigma$.  In the 1D Hubbard model ($g=0$ in
Eq.~\ref{ham}) charge and spin degrees of freedom effectively switch
places at $U=0$.  In terms of the susceptibilities, $\chi_\rho(\pi)$
and $\chi_\sigma(\pi)$ are exactly equal at $U=0$.
\begin{figure}[tb]
\centerline{\resizebox{3.2in}{!}
{\includegraphics{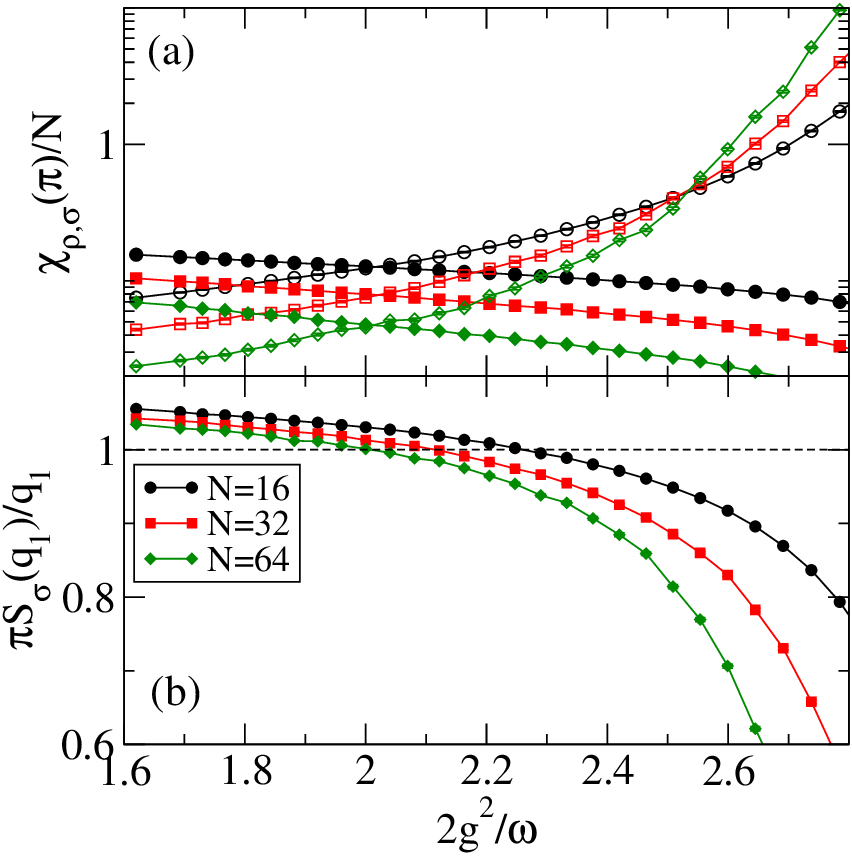}}}
\caption{(color online) (a) Charge (open symbols) and spin (filled
symbols) susceptibility for the half-filled HHM with $U=2$, $\omega=1$.
The first transition (g$_{c1}$) occurs where $\chi_\rho=\chi_\sigma$,
corresponding to $U_{\rm{eff}}=0$. The second transition (g$_{c2}$) is
the Peierls transition, where $\chi_\rho(\pi)/N$ diverges as in
Fig.~\ref{u0w1_chi_rho}.  Note that the spin susceptibility is also
divided by $N$ to make the crossing at $U_{\rm{eff}}=0$ clear.  (b)
Long-wavelength spin structure factor for $U=2$, $\omega=1$.  The
point where $\pi S_\sigma(q_1)/q_1$ crosses unity indicates the
opening of the spin gap, identical to the point where
$\chi_\rho=\chi_\sigma$ in (a).}
\label{u2w1-xrho}
\end{figure}

  In Fig.~\ref{u2w1-xrho}(a) we first show the 2k$_{\rm{F}}$ charge
and spin susceptibilities for $U=2$ and $\omega=1$.  We find that when
$U_{\rm{eff}}\approx 0$ ($g=1$ for $U=2$ and $\omega=1$), the charge
and spin susceptibilities become equal as in the simple Hubbard case.
The estimate for $K_\sigma$, shown in Fig.~\ref{u2w1-xrho}(b), again
crosses one indicating an opening of a spin gap.  This transition is
therefore the same transition $g_{c1}$ as discussed above in
Section \ref{sect:u0}, but now occurring at finite $g$.  The quantum
phase transition as $g$ increases past $g_{c1}$ appears identical to the
transition as $U$ becomes negative in the 1D Hubbard model.  Based on
the similarity with the 1D Hubbard model, we conclude that the spin
gap transition here is also of the KT form.

In Fig.~\ref{u2w1-xrho}(a) a second transition takes place beyond the
spin gap transition at $g_{c1}$. This second transition is again the
Peierls transition indicated by the divergence of
$\chi_\rho(\pi)/N$. Beyond the second transition point ($g>g_{c2}$)
$\chi_\rho(\pi)/N$ increases with increasing system size, and as in
Fig.~\ref{u0w1_chi_rho}(b) $\chi_\rho(\pi)/N$ for different system
sizes cross at $g=g_{c2}$ when plotted versus e-ph coupling.  For
$g_{c1}<g<g_{c2}$ we now have a third intermediate phase, which has a
spin gap but no Peierls order.  In Fig.~\ref{u2w1-xrho}(a) we see only
very small finite-size effects in determining $g_{c1}$ and $g_{c2}$
from the susceptibility data.  The $g_{c1}$ from our data shows little
deviation from $U_{\rm{eff}}=0$, at least for small to intermediate
$U$ as compared to $\omega$.  Finite-size effects are more significant
in $K_\sigma$ as estimated from the spin structure factor slope in
Fig.~\ref{u2w1-xrho}(b) because $q_1=2\pi/N$ only approaches $q_1=0$
in the limit $N\rightarrow \infty$.  However, for increasing $N$,
$g_{c1}$ as estimated from $K_\sigma$ does converge to the same value
we obtain from the susceptibility. 

 As $U$ increases, we find that two transitions at $g_{c1}$ and
$g_{c2}$ occur closer together, becoming indistinguishable from each
other at approximately $U\sim 5$ for $\omega=1$. At this point and for
larger $U$, the two KT transitions merge into a single Mott/Peierls
transition. We next show that this merged transition is {\it first
order}.

\subsection{First order transition}

\begin{figure}[tb]
\centerline{\resizebox{3.0in}{!}
{\includegraphics{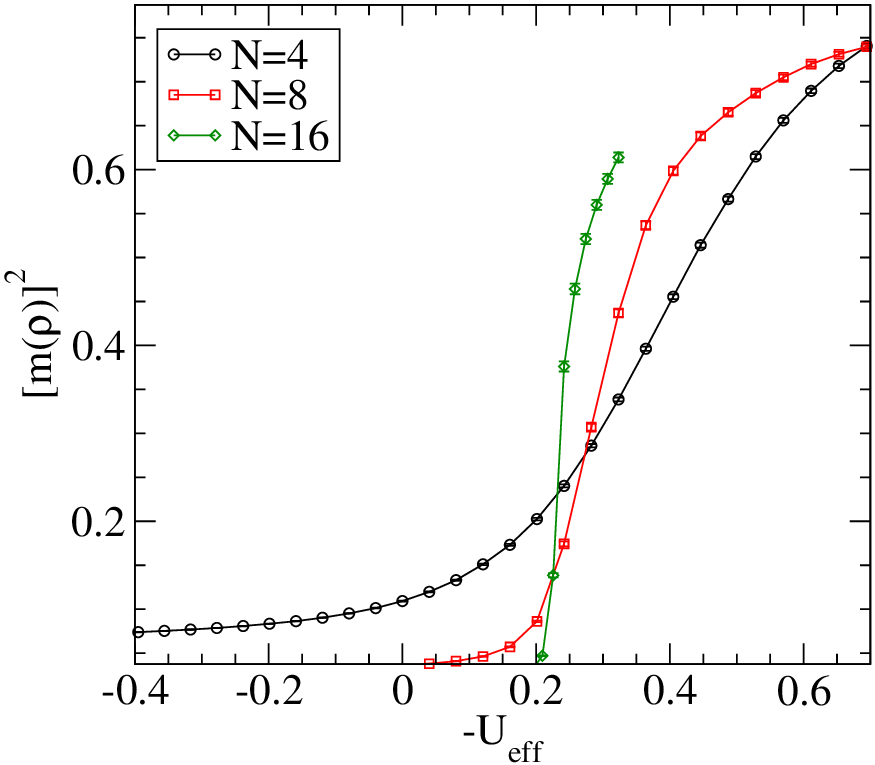}}}
\caption{(color online) First order Mott/Peierls transition for large
$U$. We show the CDW order parameter $[m(\rho)]^2$ (see text)
vs. $-U_{eff}=2g^2/\omega-U$ for $\omega=1$ and $U=8$.  The
Mott/Peierls transition occurs for $U_{\rm{eff}}\sim -0.2$.}
\label{u8w1-firstorder}
\end{figure}

Above a critical $U$ value $U=U_m$ we find that the spin-gap and
Peierls transitions coincide. The phase diagram then has a shape very
similar to that of the half-filled 1D extended Hubbard model (EHM)
\cite{Hirsch83b,Cannon90a,Cannon91a,Sengupta02a}. In the half-filled
EHM, as the nearest-neighbor Coulomb repulsion $V$ is increased for
fixed $U$, there is a transition from AFM to CDW order. This
transition is continuous for small $U$ and first-order for $U>U_m$.
In a first-order quantum phase transition, observables become
discontinuous as one of the Hamiltonian parameters is varied.  For the
HHM, a change to first order behavior for strong coupling has also been
seen in DMFT studies\cite{Koller04a}.  As in Reference
\onlinecite{Sengupta02a} we take $[m(\rho)]^2=S_\rho(\pi)/N$ as an
order parameter for the Peierls CDW state.  We show in
Fig.~\ref{u8w1-firstorder} $[m(\rho)]^2$ for $U=8$ and $\omega=1$.  We
find a sharp jump in $[m(\rho)]^2$ at the transition with the
discontinuity becoming stronger for larger system sizes. Other
observables such as the ground state energy and bond order also show
discontinuous behavior consistent with a first order transition.  In
fact, this point is a {\it multi-critical} point. In the EHM there is
an intervening phase with long-range BOW for $U<U_m$
\cite{Nakamura00a,Nakamura00b,Sengupta02a}. We find very similar
behavior in the HHM except that the intervening phase here is the
metallic intermediate state.  We cannot calculate a precise value for
$U_m$, but for $\omega=1$ it appears comparable ($U_m\sim 5$ for
$\omega=1$) to the value found in the half-filled EHM, $U_m=4.7\pm 0.1$
\cite{Sengupta02a}. We also remark that the change in the order of the
transition may be related to discussions of quantum to classical
crossover in e-ph coupled models \cite{Caron84a}.

\subsection{Discussion of Luttinger exponents}
\label{sect:krho}

\begin{figure}[tb]
\centerline{\resizebox{3.2in}{!}
{\includegraphics{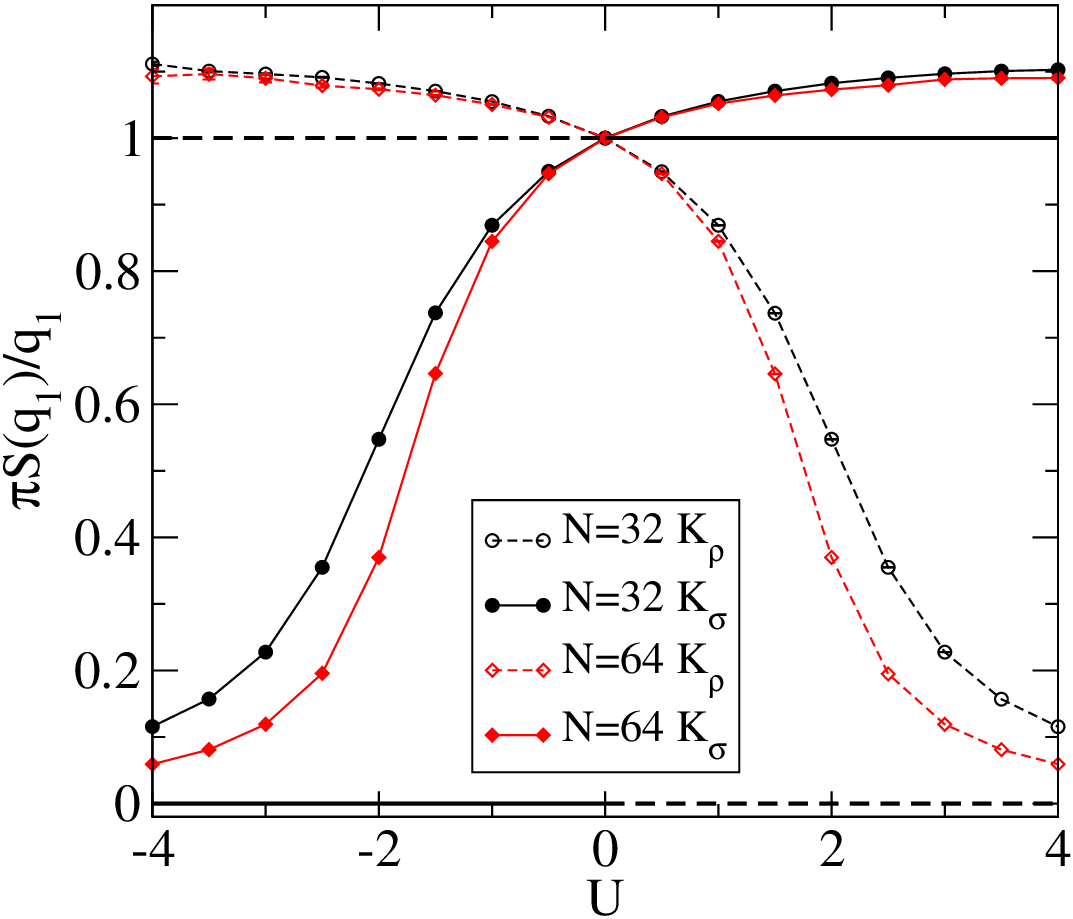}}}
\caption{(color online) LL exponents for the 1D Hubbard model
(Eq.~\ref{ham} with $g=0$) estimated from the long-wavelength charge
and spin correlations.  $K_\rho$ ($K_\sigma$) is given by open
(filled) symbols.  In the infinite $N$ limit, $K_\rho=1$ for any $U<0$
and $K_\rho=0$ for $U>0$; $K_\sigma=1$ for any $U>0$ and $K_\sigma=0$
for any $U<0$. Observing these limiting values (shown by full and
dashed horizontal lines) is difficult due to finite value of
$q_1=2\pi/N$ and also the logarithmic scaling with $N$ for the
exponent whose value is unity.}
\label{1dhubbard}
\end{figure}
In the LL picture, $K_\rho$ and $K_\sigma$ determine the asymptotic
decay of correlations functions and hence measurements of these
exponents in finite systems have often been used to determine the
phase diagrams of 1D models.  Specifically, $K_\rho>1$ corresponds to
attractive charge correlations, while $K_\rho<1$ corresponds to
repulsive charge correlations.  It is first instructive to review the
LL exponents for the 1D Hubbard model and sources of error in
finite-size systems.  At half filling for $U>0$ the 1D Hubbard model is
insulating ($K_\rho=0$) with no spin gap ($K_\sigma=1$, spin
rotational invariance holds).  For $U<0$ there is a spin gap
($K_\sigma=0$), and degenerate CDW and singlet superconducting (SS)
pair correlations ($K_\rho=1$). Therefore, the LL exponents are {\it
discontinuous} at $U=0$.  The transition at $U=0$ is of the KT type,
with the gaps (charge gap $U>0$ or spin gap $U<0$) opening
exponentially slowly as $U$ is varied from zero. In
Fig.~\ref{1dhubbard} we show $K_\rho$ and $K_\sigma$ for the 1D
Hubbard model calculated using Eq.~\ref{krho}.  There are two primary
sources of finite-size error: first, the requirement that
$q\rightarrow 0$ in Eq.~\ref{krho}, and second, the presence of
logarithmic scaling corrections near a KT transition.  The scaling
with system size is slow close to the transition ($U=0$) and
particularly slow for the exponent that is expected to be equal to 1
($K_\sigma$ for $U>0$ and $K_\rho$ for $U<0$). Such logarithmic
scaling has been noted in other 1D electron and spin models and makes
it difficult in practice to observe $K_\sigma=1$ for the positive-$U$
Hubbard model in a finite-size calculation
\cite{Sandvik99b,Sengupta03a}.  As discussed in Section \ref{sect:u0},
log corrections are expected to vanish exactly at critical coupling.
In Fig.~\ref{1dhubbard} this occurs at $U=0$, where $K_\rho$ and
$K_\sigma$ curves for all system sizes cross at $K_\rho=K_\sigma=1$.

Turning now to the HHM, the variation of $K_\sigma$ for $g<g_{c1}$
(Fig.~\ref{u2w1-xrho}(b)) is consistent with log corrections in the
spin degree of freedom that vanish at the spin gap transition. This
observation further reinforces our statement that the spin gap
transition is also of the KT type.  For the $K_\rho$ data in
Fig.~\ref{u0w1_ll}, $K_\rho$ at $g=0$ is again exactly unity. $K_\rho$
then crosses one from above at a $g$ roughly consistent with the
$g_{c2}$ determined from the susceptibility data in
Fig.~\ref{u0w1_chi_rho}. Assuming the Peierls transition occurs where
$K_\rho=1$ gives a critical coupling of $2g^2_{c2}\omega\approx 0.85$
after performing finite-size scaling using $N$ up to 128 sites
(Fig.~\ref{u0w1_ll}).  The form of the $K_\rho$ plot for the HHM
($U=0$) is clearly similar to $K_\rho$ for the negative $U$ Hubbard
model (Fig.~\ref{1dhubbard}), with $K_\rho$ starting at one for zero
coupling, and becoming slightly {\it larger} than one for nonzero
coupling. While this apparent $K_\rho>1$ may be interpreted as meaning
that superconducting pair correlations are dominant\cite{Clay05b}, a
more plausible interpretation is that the apparent $K_\rho>1$ is a
consequence of logarithmic scaling corrections.  This implies that the
true $K_\rho$ {\it should be exactly equal to unity} for $g<g_{c2}$,
and drop to zero for $g>g_{c2}$.  This further implies that the
intermediate state has {\it degenerate} CDW and SS correlations.  This
statement is consistent with our finding that the $U=0$ HHM for
$g<g_{c2}$ has a spin gap but no charge gap.

Calculations for a model of acoustic phonons coupled to 1D electrons
found that the LL expressions for decay of correlation functions must
be modified due to retardation effects\cite{Loss94a}. Specifically,
the dominance of CDW and SS correlations is given by
\begin{eqnarray}
K_\rho A & \leq & 1 \qquad \rm(CDW) \label{krho-renorm}\\
B/K_\rho  &\leq& 1 \qquad (SS) \label{krho-renorm-ss}
\end{eqnarray}
where $A$ and $B$ depend on the strength of the e-ph coupling
\cite{Loss94a}. With zero e-ph coupling, $A=B=1$.  For increasing e-ph
coupling, $A>1$ and $B<1$, with $A$ diverging and $B$ approaching a
finite value.  The renormalized boundary for the metallic/Peierls
transition is then $K_\rho=1/A$.  While there is no reason to expect
that for the HHM model (with dispersionless phonons) the LL relations
should be renormalized in the same manner, our SSE data may be
consistent with $1/A$ slightly less than 1. Upon close examination of
Fig.~\ref{u0w1_ll} and Fig.~\ref{u0w1_chi_rho}, the $g_{c2}$ as
determined by $K_\rho$ crossing one is slightly smaller than
the $g_{c2}$ determined by susceptibility.  The $g_{c2}$ determined
from $K_\rho$ (Fig.~\ref{u0w1_ll}) would coincide with the $g_{c2}$
determined from $\chi_\rho(\pi)$ (Fig.~\ref{u0w1_chi_rho}(b)) if the
horizontal line in Fig.~\ref{u0w1_ll} is moved slightly below one, or
$1/A\approx 0.95$.

For larger $U$, the size of the intermediate region shrinks, and
$K_\rho$ peaks at the transition, with $K_\rho$ approaching one with
increasing $N$ (see Fig.~2(a) in Reference \onlinecite{Clay05b},
$U=2$, $\omega=0.5$). The peak at the transition is consistent with
$K_\rho=0$ in the Mott and Peierls states, and $K_\rho=1$ only along
their boundary. The apparent $K_\rho<1$ at the peak may be due to the
closer proximity to the first order transition, where $K_\rho$ drops
quite rapidly to zero. For $U=2$ and $\omega=0.5$, we estimate that
$0.95\alt 1/A \leq 1$.  
If renormalization as in Reference \onlinecite{Loss94a} does occur, for
all parameter values we investigated it appears that the effect is
relatively small ($0.9\alt A \alt 1$).
Because measuring SS correlations is not practical in the SSE method,
we cannot determine a value for $B$. Eq.~\ref{krho-renorm-ss} with
$B<1$ would imply that a SS correlations are dominant whenever
$K_\rho$ exceeds a value that is {\it smaller} than one. SS is
dominant for any nonzero e-ph coupling for $U=0$ in the calculation of
Reference \onlinecite{Loss94a}, which seems unlikely in the HHM. We
will discuss these implications further in Section
\ref{sect:conclusion}.

\subsection{Phase diagram, half filling}

\begin{figure}[t]
\centerline{\resizebox{3.2in}{!}{\includegraphics{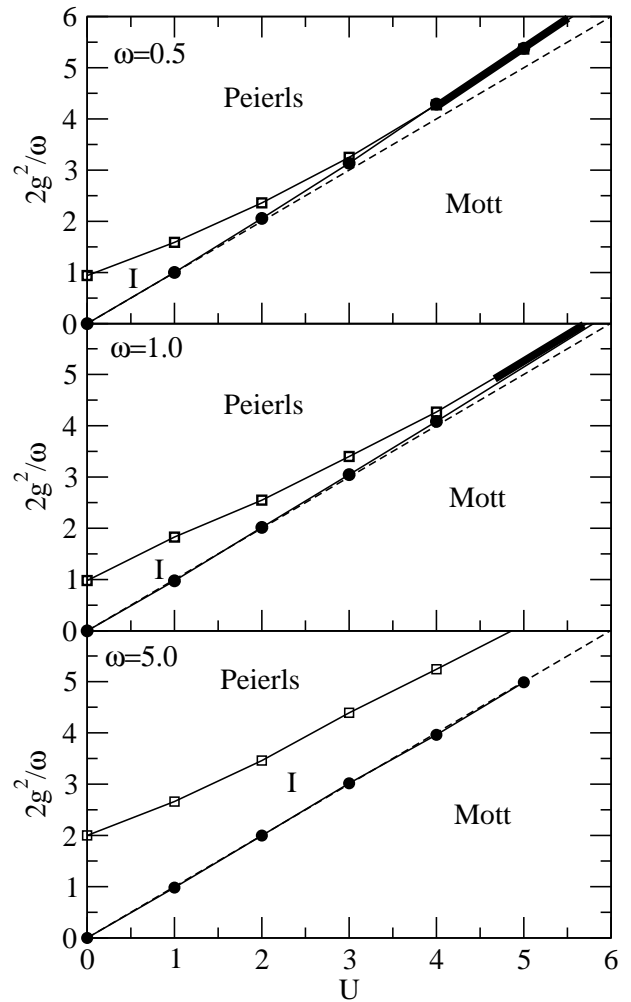}}}
\caption{Phase diagram of the half-filled HHM for $\omega=0.5$,
$\omega=1$, and $\omega=5$. The dashed line is given by
$U=2g^2/\omega$.  All phase boundaries are determined using
susceptibility and $K_\sigma$ data, with uncertainty approximately the
size of the symbols. Lines are guides to the eye.  The three phases
shown are Mott, (I)ntermediate, and Peierls. The Mott/I and I/Peierls
boundaries merge into a single first-order Mott/Peierls boundary
indicated by a heavy line for $U\agt 4$ for $\omega=0.5$ and $U\agt 5$
for $\omega=1$.}
\label{half-phasediag}
\end{figure}
In Fig.~\ref{half-phasediag} we show the phase diagram for
$\omega=0.5$, $\omega=1$, and $\omega=5$. All points were determined
using susceptibility data for systems up to 32 (and in some cases 64
and 128) sites. We find that with increasing $\omega$ the width of the
intermediate region increases, and the tricritical point $U_m$ moves
to larger $U$.  One further observation is that for $U\agt U_m$, the
deviation of the Mott/Peierls boundary from $U_{\rm{eff}}=0$ becomes
noticeable, with the boundary shifting to $U_{\rm{eff}}<0$ (above the
dotted lines in Fig.~\ref{half-phasediag}).  This shift can be seen for
example in Fig.~\ref{u8w1-firstorder}. For $U<U_m$, the
Mott/intermediate spin gap boundary is very close to the line
$U_{\rm{eff}}=0$.

\section{Quarter filling}

\begin{figure}[tb]
\centerline{\resizebox{3.2in}{!}
{\includegraphics{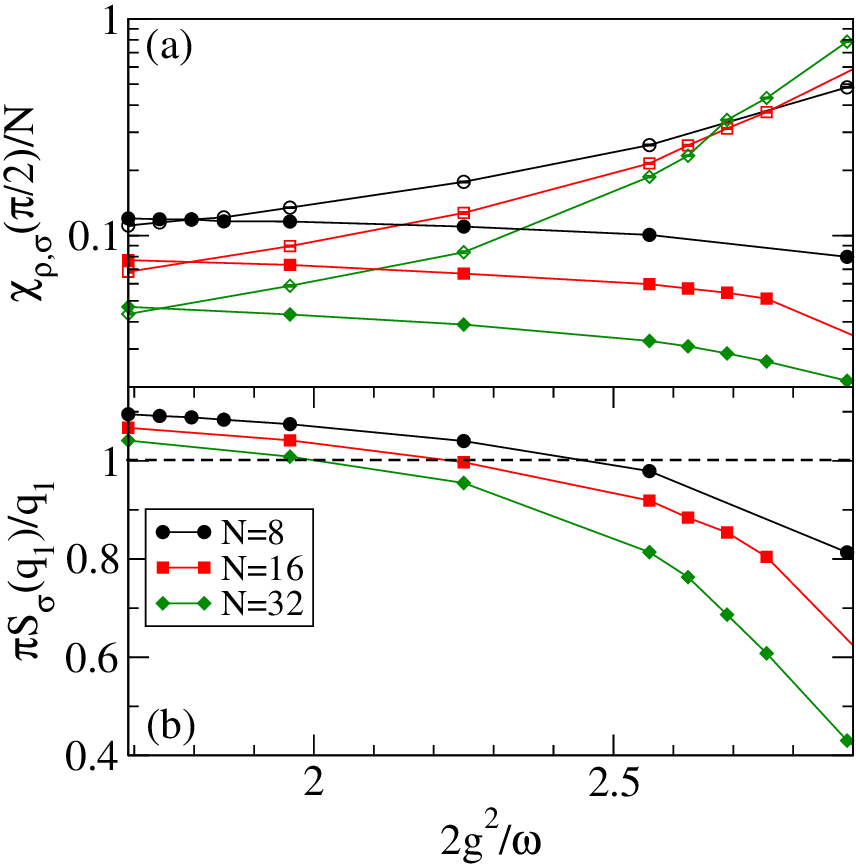}}}
\caption{(color online) (a) Charge (open symbols) and spin (filled symbols)
susceptibilities for the quarter-filled HHM with $U=2$,
$\omega=0.5$. (b) Long-wavelength spin structure factor for same
parameters. We find similar behavior to half filling,
Fig.~\ref{u2w1-xrho}, with first a transition to a spin-gapped state,
and second the transition to the Peierls CDW state.}
\label{u1w1-qtr}
\end{figure}
Many of the materials that the HHM is most applicable to are not
half-filled. For example, most of the quasi-1D organic superconductors
are 3/4 filled (1/4 hole filled) \cite{Ishiguro}.  We therefore
present some results for the HHM at quarter filling.  Although for
many of these materials it is necessary to include long ranged Coulomb
interactions (the extended Hubbard $V$ term) \cite{Hubbard78a}, we
will continue to focus on the HHM Hamiltonian with only onsite $U$ and
e-ph terms. We comment on the expected effect of $V$ further below. As
quarter filling is commensurate a Peierls state is also expected to
occur for sufficiently large $g$. There are however significant
differences between half-filled and quarter-filled Peierls states.  At
quarter filling there are more than one possible pattern of charge and
bond distortion, and which one actually occurs depends on the values
of $U$ as well as $V$ \cite{Ung94a,Clay03a}.  In the absence of
phonons, the quarter-filled band for finite $U$ is a LL with neither
charge nor spin gaps. As at half filling, $\chi_\rho(2k_{\rm{F}})$ and
$\chi_\sigma(2k_{\rm{F}})$ are degenerate at $U=0$ (note that
2k$_{\rm{F}}=\pi/2$ at quarter filling and corresponds to a
correlation function with period 4 in real space).  In the presence of
phonons, we again expect the charge susceptibility
$\chi_\rho(2k_{\rm{F}})/N$ to diverge.

Our SSE results show that the HHM at quarter filling is in many respects
similar to the half-filled case. In Fig.~\ref{u1w1-qtr} we show
$\chi_\rho(2k_{\rm{F}})/N$ and $\pi S_\sigma(q_1)/q_1$ versus
$2g^2/\omega$. We again find two transitions: first a transition to a
spin-gapped state, and second the transition to the Peierls state. As
at half filling, the spin gap opens very close to the point where
$U_{\rm{eff}}=U-2g^2/\omega=0$. The phase diagram at quarter filling is
therefore nearly identical to the phase diagram at half filling, with
LL, intermediate, and Peierls phases.  We find that the intermediate
phase is slightly wider at quarter than half filling. For example, at quarter
filling with $U=2$ and $\omega=0.5$ (Fig.~\ref{u1w1-qtr}),
$2g^2_{c1}/\omega\approx 1.7$ and $2g^2_{c2}/\omega\approx 2.6$,
compared to $2g^2_{c1}/\omega\approx 2.0$ and $2g^2_{c2}/\omega\approx
2.3$ at half filling. We note (see Fig.~\ref{u1w1-qtr}) that at quarter
filling we see slightly greater deviation from the $U_{\rm{eff}}=0$ in
the first (spin gap) transition.  At present we do not have enough SSE
data to investigate whether the tricritical point $U_m$ occurs as at
half filling, but in our data at quarter filling we do find that with
increasing $U$ $g_{c1}$ and $g_{c2}$ become closer together. This
suggests that a tricritical point also exists at quarter filling.

\begin{figure}[tb]
\centerline{\resizebox{3.0in}{!}
{\includegraphics{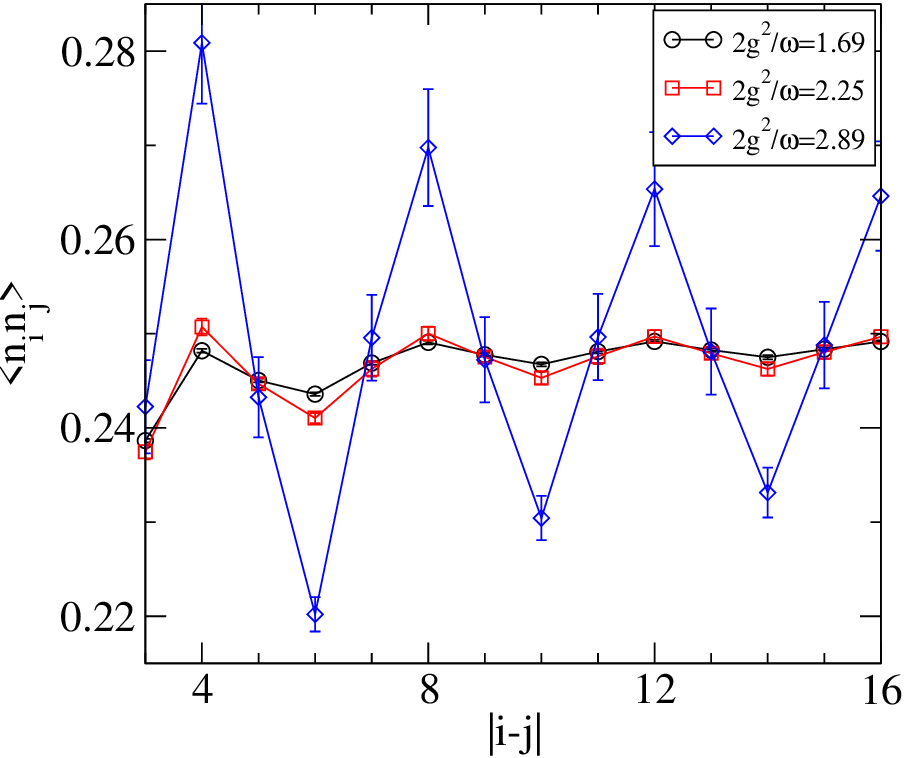}}}
\caption{(color online) Charge-charge correlations $\langle n_i n_j
\rangle$ versus distance $|i-j|$ for a 32 site quarter-filled system with
$U=2$ and $\omega=0.5$.  The three values of $2g^2/\omega=1.69$, 2.25,
and 2.89 correspond to LL, intermediate, and Peierls states,
respectively. We find that in all three regions the charge
correlations at quarter filling are of the form $\cdots$2000$\cdots$.}
\label{qtr-chgchg}
\end{figure}
At quarter filling there are two possible CDW's that are period-4
(2k$_F$). These have charge densities in cartoon form of either
$\cdots$1100$\cdots$ or $\cdots$2000$\cdots$, where ``1'' or ``2''
indicates a charge density greater than the average density of 0.5 and
``0'' indicates a charge density less than the average \cite{Ung94a}.
The pattern $\cdots$2000$\cdots$ is found in the uncorrelated ($U=0$)
band.  In Fig.~\ref{qtr-chgchg} we plot the real-space charge-charge
correlation function $\langle n_i n_j \rangle$ versus distance $|i-j|$
for a range of $g$'s in the three phases. We find that the
charge-charge correlation function peaks for sites separated by four
lattice sites, consistent with a CDW state of the $\cdots$2000$\cdots$
form. The strength of the CDW correlations does not greatly change
going from the LL to the intermediate phase, but increases rapidly
after the Peierls transition. In the $\cdots$2000$\cdots$ CDW, the
three small charges are not exactly equal, and the actual charge
densities are in sequence large, medium, small, medium (LMSM).  This
charge pattern coexists with a BOW because L-M and M-S bonds are
inequivalent.  Fig.~\ref{qtr-chgchg} shows that the charge
correlations follow this LMSM pattern as expected. We conclude that
the pairing at quarter filling in the HHM consists of onsite electron
pairs as found at half filling, at least for the small through
intermediate $U$ we have currently investigated.

The distinction between these two CDW patterns at quarter filling is
important because while $\cdots$2000$\cdots$ is related to {\it
onsite} electron pairs, the more extended CDW $\cdots$1100$\cdots$ is
related to {\it nearest-neighbor} pairing. The $\cdots$1100$\cdots$
requires bond-coupled phonons in addition to the Holstein phonons
considered here \cite{Ung94a,Clay03a}. In addition, the pattern of the
BOW (the location of the ``strong'' bond) coexisting with the
$\cdots$1100$\cdots$ CDW also depends on the strength of $V$
\cite{Ung94a}. If a similar metallic phase exists adjacent to the
$\cdots$1100$\cdots$ CDW, it is possible that a region of
nearest-neighbor superconducting pairing will be found that may be
relevant to real quarter-filled molecular superconductors.

\section{Conclusions}
\label{sect:conclusion}

To summarize, we have presented numerical data for charge and spin
correlations of the 1D HHM model at half and quarter filling.  We have
based our phase diagram on charge and spin susceptibilities, which
provide direct indication of phase boundaries with much weaker
finite-size effects than previous calculations based on LL exponents
\cite{Clay05b}.  We find that the spin gap and Peierls transitions do
not occur simultaneously unless $U$ is larger than a critical
$U_m$. For $U<U_m$ as the e-ph coupling is increased from zero, the
spin gap opens {\it before} the Peierls state forms. The intermediate
state is metallic with a spin gap but no charge gap, and the
transitions to and from the intermediate state are of the KT type.
Our physical picture of the intermediate state is that at the spin gap
transition ($g_{c1}$), pairs are formed, but are disordered
and do not order in a Peierls state until the e-ph coupling is further
increased.  For $U>U_m$, the two transitions merge into a single
first-order Mott/Peierls transition. With finite-size calculations we
cannot completely discount the possibility of a small charge gap
(small compared to the finite-size gap) in the intermediate
region. However, finite charge stiffness (Drude weight) provides further
evidence for metallic behavior in the intermediate state
\cite{Clay05b}.

Comparing to other calculations, the critical coupling we determined
for $g_{c2}$ at $U=0$ is consistent with previous results
\cite{Wu95a,Jeckelmann99a}. The variational results of Reference
\onlinecite{Takada03a} find the intermediate phase existing in a
narrow region on both sides of the $U_{\rm{eff}}=0$ line, while we
find the intermediate phase only for $U_{\rm{eff}}<0$.  Several 
calculations of the single-particle spectral function are available
for the HHM \cite{Fehske04a,Ning06a,Matsueda06a}, the spinless Holstein model
\cite{Hohenadler06a,Sykora06a}, as well as the $d=\infty$ studies
previously mentioned. In Reference \cite{Ning06a} using a
cluster perturbation theory method applied to the 1D HHM, a small
nearly dispersionless peak was found in the spectral function for
small $k$. This small peak is also found in the spectral function of
the metallic phase of the spinless Holstein model and may possibly
be associated with the intermediate phase.

Considering the possible modification of the LL equations in the
presence of retarded e-ph interactions, we find that while this could
possibly occur in a form that would agree with Reference
\onlinecite{Loss94a}, the amount of renormalization is small ($1/A\sim
0.95$), and possibly within finite-size errors in our determination of
the transition points. We also do not see any measurable or consistent
change in the constant $A$ when comparing $\omega=0.5$ and $\omega=1$,
which would be expected to change the amount of retardation in the
e-ph interaction.  We are not able to calculate a value for $B$ in
Eq.~\ref{krho-renorm-ss}.  However, if Eq.~\ref{krho-renorm-ss} is
correct for the HHM model with $B<1$, SS correlations would actually
be {\it enhanced} because of retardation \cite{Loss94a}.

An important question is the strength of SS correlations within the
intermediate region. In terms of the LL framework, models with
$K_\rho>1$ have dominant superconducting correlations. Indeed, our
numerical data appears to show $K_\rho>1$ in the intermediate region,
but this result is likely to be a finite-size effect.  If we set aside
any renormalization of $K_\rho$, our conclusion based on comparison
with the 1D negative-$U$ Hubbard model is that in the intermediate
region, $K_\rho$ is {\it exactly} equal to one.  This implies that in
the intermediate region that CDW and SS correlations are in fact {\it
exactly degenerate}.  This exact degeneracy may not be easily
observable in a finite system due to the finite size difficulties near
KT transitions. As the SSE method is based on a world-line approach in
imaginary time, there is no simple way to measure correlations
involving four particles, which would be needed to measure SS
correlations directly. Other QMC \cite{Tam06a} and DMRG
\cite{Tezuka05a} calculations suggest that while SS and CDW
correlations are nearly degenerate in the intermediate region, CDW
correlations appear slightly stronger at long range.  DMRG results for
a ladder system suggest that going beyond 1D can break the degeneracy,
giving a region with dominant SS correlations \cite{Tezuka05a}.
Finally, we remark that in addition to the logarithmic scaling
difficulties, observing metallic behavior in close proximity to a CDW
is difficult due to the typically rapidly increasing autocorrelation
time for QMC methods.  We do find in our method that the
autocorrelation time $\tau$ increases rapidly close to the Peierls
boundary, but believe that the use of parallel tempering can reduce
$\tau$ enough to obtain reliable results.

The authors acknowledge support of American Chemical Society Petroleum
Research Fund, and the Department of Energy grant DE-FG02-06ER46315.

\end{document}